\documentclass[10pt,letterpaper]{article}
\usepackage{opex3}
\usepackage{hyperref}

\usepackage{hyperref}  
\usepackage{eqnarray}
\usepackage{amsmath}
\usepackage{stackrel}

\usepackage{color}

\definecolor{mygreen}{rgb}{0,0.5,0} 
\definecolor{myblue}{rgb}{0,0,0.75} 
\definecolor{myyellow}{rgb}{0.87,0.8,0.47} 
\definecolor{mymagenta}{cmyk}{0,1,0,0.12}

\begin{document}

\title{Atomic filtering for hybrid continuous-variable/discrete-variable quantum optics}

\author{Joanna A. Zieli\'nska,$^{1,*}$ Federica A. Beduini,$^{1}$ Vito Giovanni Lucivero,$^{1}$ and Morgan W. Mitchell$^{1,2}$}

\address{$^1$ICFO-Institut de Ciencies Fotoniques, Av. Carl Friedrich Gauss, 3, 08860 Castelldefels, Barcelona, Spain\\
$^2$ICREA-Instituci\'{o} Catalana de Recerca i Estudis Avan\c{c}ats, 08015 Barcelona, Spain}

\email{$^*$joanna.zielinska@icfo.es} 

\begin{abstract}

We demonstrate atomic filtering of frequency-degenerate photon pairs from a sub-threshold optical parametric oscillator (OPO).  The filter, a modified Faraday anomalous dispersion optical filter (FADOF), achieves 70\% peak transmission simultaneous with 57 dB out-of-band rejection and a 445 MHz transmission bandwidth.  When applied to the OPO output, only the degenerate mode, containing one-mode squeezed vacuum, falls in the filter pass-band; all other modes are strongly suppressed.  The high transmission preserves non-classical continuous-variable features, e.g. squeezing or non-gaussianity, while the high spectral purity allows reliable discrete-variable detection and heralding.  
Correlation and atomic absorption measurements indicate a spectral purity of 96\% for the individual photons, and 98\% for the photon pairs.  These capabilities will enable generation of atom-resonant hybrid states, e.g. ``Schr\"odinger kittens'' obtained by photon subtraction from squeezed vacuum, making these exotic states available for quantum networking and atomic quantum metrology applications.  
\end{abstract}



\bibliography{BigBibMWM,FDC_bib}
\bibliographystyle{osajnl}

\section{Introduction}

Experimental and theoretical methods for studying quantum fields have traditionally been divided between the ``continous-variable'' and ``discrete-variable'' camps, each with distinct language and experimental techniques \cite{BraunsteinBOOK2010}.  Recently this artificial division has begun to dissolve, and experiments combining continuous-variable and discrete-variable elements \cite{LvovskyPRL2001, OurjoumtsevS2006, Neergaard-NielsenPRL2006, DeMartiniPRL2008, LvovskyNPhys2013,BrunoNPhys2013, MorinARX2013, AndersenPRA2013, JeongARX2013} have proliferated.  These hybrid methods create new possibilities, including  optical entanglement between particle-like and wave-like states \cite{MorinARX2013, AndersenPRA2013, JeongARX2013}, a form of micro-macro entanglement \cite{DeMartiniPRL2008, LvovskyNPhys2013,BrunoNPhys2013} that puts to test conventional notions of the quantum/classical boundary.
Hybrid approaches have been proposed for loophole-free Bell inequality tests \cite{GarciaPatronPRL2004,AcinPRA2009}, quantum metrology \cite{JooPRL2011,WolfgrammNPhot2013} and quantum computing \cite{GilchristJOB2004}.  

Interfacing hybrid states to single atoms or atomic ensembles would further expand their power, offering synchronization in communications and computing protocols \cite{BriegelPRL1998}, quantum-enhanced probing of atomic sensors \cite{WolfgrammPRL2010, WolfgrammNPhot2013}, and tests of quantum non-locality with massive particles \cite{RoweN2001}.  If optical hybrid states can be transferred to an atomic system, they can be detected by quantum non-demolition measurement \cite{KoschorreckPRL2010a,KoschorreckPRL2010b,DubostPRL2012, ChristensenNJP2013, SewellNP2013, ChristensenPRA2014}, allowing non-destructive characterization and repeated use.  A major challenge for the interaction of non-classical states with atomic systems has been generating quantum light at the wavelengths and bandwidths of atomic transitions \cite{WolfgrammOE2008, PredojevicPRA2008, Predojevic2009, RielanderPRL2014}.  The hybrid continous-discrete variable approach offers still more challenges: the heralding process must be highly selective to avoid false heralding events, while the continuous-variable states must be protected against both dephasing and loss.   

Here we present an essential step toward the use of hybrid optical states with atomic systems, by demonstrating the efficient isolation of atom-resonant single-mode squeezed vacuum, the starting ingredient for production of ``Schr\"{o}dinger kitten'' states \cite{OurjoumtsevS2006},  from a much stronger broadband background of two-mode squeezed states, the natural output of a sub-threshold OPO or cavity-enhanced SPDC source.  
We work at the D$_1$ line of atomic rubidium at 794.7 nm, a favorite wavelength for atomic quantum memories.  Thulium-doped solid state quantum memories \cite{SaglamyurekN2011, PascualWinterPRB2012} also operate at nearly this same wavelength.  The squeezed vacuum is generated by a sub-threshold OPO consisting of an optical resonator with a blue-light-pumped $\chi^{(2)}$ nonlinear medium inside.  The OPO generates narrowband near atom-resonant squeezed vacuum in the degenerate cavity mode, i.e., the longitudinal mode with half the pump frequency, but also a far larger number of two-mode squeezed states in other modes.  We use a modified Faraday rotation anomalous-dispersion optical filter (FADOF) \cite{zielinskaOL2012} to separate the single-mode squeezed vacuum from these other, co-propagating, modes.  Previously, atomic filters have been used to filter single photons \cite{WolfgrammPRL2011} and polarization-distinguishable photon pairs \cite{WolfgrammNPhot2013}, but with lower efficiencies (up to 14\%), incompatible with non-classical continuous-variable states. 

With this new FADOF we observe 70\% transmission of the degenerate mode through the filter, compatible with 5 dB of squeezing, simultaneous with out-of-band rejection by 57 dB, sufficient to reduce the combined non-degenerate emission to a small fraction of the desired, degenerate mode emission.   In comparison, a recently-described monolithic filter cavity achieved 60\% transmission and 45 dB out-of-band rejection \cite{PalittapongarnpimRSI2012}.  We test the filter by coincidence detection of photon pairs from the squeezed vacuum, which provides a stringent test of the suitability for use at the single-photon level.  We observe for the first time fully-degenerate, near atom-resonant photon pairs, as evidenced by correlation functions and  atomic absorption measurements.  The 96\% spectral purity we observe is the highest yet reported for photon pairs, surpassing the previous record of 94\% \cite{WolfgrammPRL2011}, and in agreement with theoretical predictions.  
  
\section{Experiment}
\label{sec:Experiment}
\subsection{Source of photon pairs}

In our experiment, a doubly-resonant degenerate OPO \cite{Anacavity} featuring a type-I PPKTP crystal produces single-mode squeezed vacuum at 794.7 nm. A continuous wave external cavity diode laser is stabilised at the frequency $\omega_0$ of maximum transmission of the FADOF (2.7 GHz to the red of the Rb $D_1$ line centre, as in \cite{zielinskaOL2012}): an electro-optic modulator (EOM) adds sidebands to the saturated spectroscopy absorption signal in order to get an error signal at the right frequency. In order to generate pairs at $\omega_0$, we double the laser frequency, via cavity-enhanced second harmonic generation in a LBO crystal, generating the 397.4 nm pump beam for the OPO.

With this configuration, photon pairs are generated at the resonance frequencies of the OPO cavity that fall inside the  150~GHz-wide phase matching envelope of the PPKTP crystal. Hence, the OPO output is composed by hundreds of frequency modes, each of 8.4~MHz bandwidth, separated by the 501~MHz free spectral range. This means that the FADOF - with its 445~MHz bandwidth - can successfully filter all the nondegenerate modes, leaving only the photons in the degenerate mode, which are then fully indistinguishable, as they share the same spatial mode, frequency and polarization.

The light generated by the OPO is then sent through a polarization-maintaining fiber to the filter setup, filtered by the FADOF described in the next subsection, and coupled to a fiber beamsplitter. The detection scheme is described in the subsection \ref{sec:detection}.
\begin{figure}[htb]
 \centerline{\includegraphics[width=14cm]{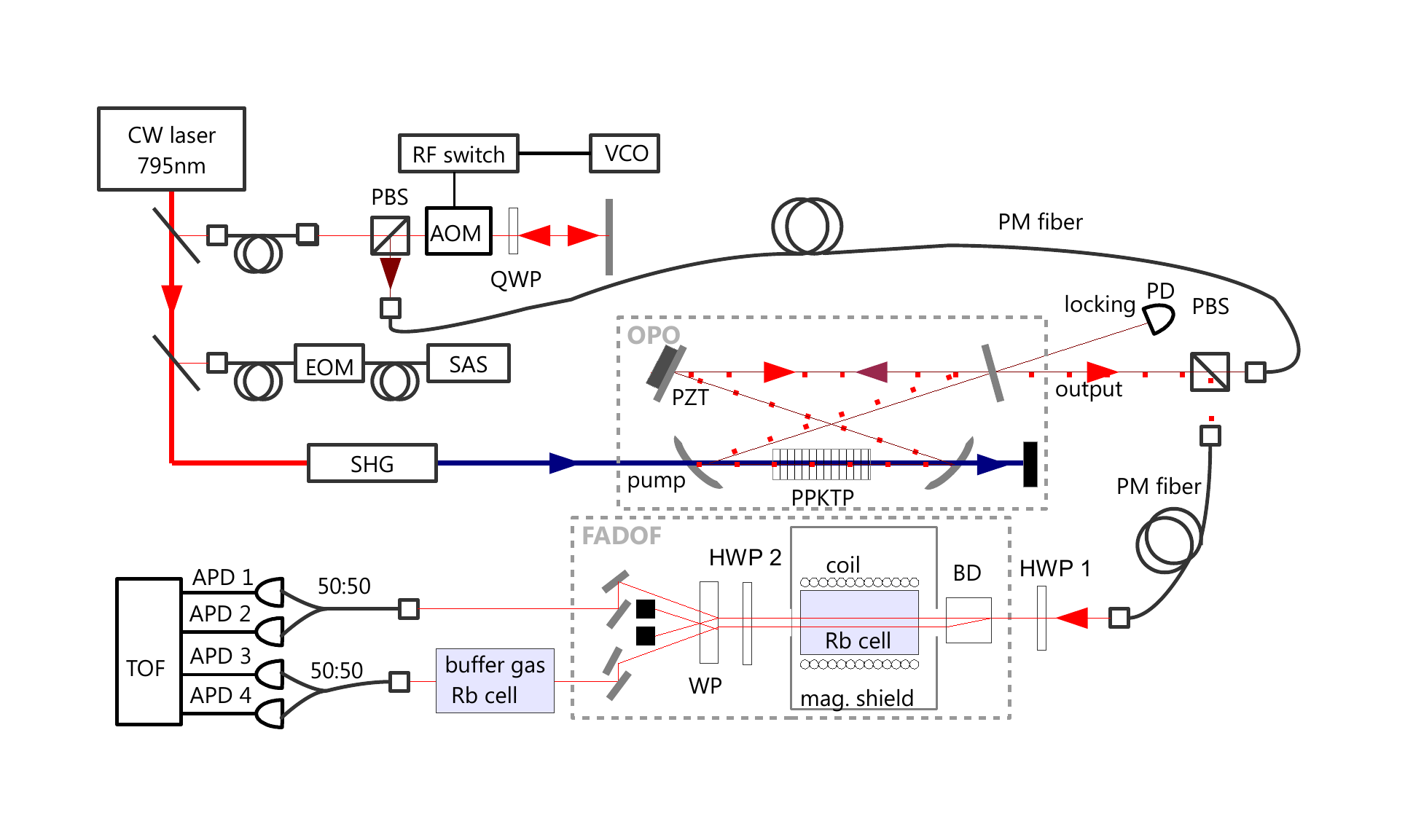}}
\caption{\label{fig:opo}Experimental setup of the OPO, the FADOF filter and detection system. Symbols: PBS: polarizing beam splitter, AOM: acousto-optic modulator, EOM: electro-optic modulator, APD: avalanche photodiode, BD: calcite beam displacer, WP: Wollaston prism, TOF: time-of-flight analyzer, SAS: saturated absorption spectroscopy, VCO: voltage controlled oscillator, PM: polarization maintaining fiber, HWP: half-wave plate, QWP: quarter-wave plate, PD: photodiode}
\end{figure}

\subsection{Faraday Anomalous Dispersion Optical Filter}
 
The FADOF consists of a hot atomic vapor cell between two crossed polarizers that block transmission away from the absorption line, while the absorption itself blocks resonant light. A homogeneous magnetic field along the propagation direction induces circular birefringence in the vapor, so that the Faraday rotation just outside the Doppler profiles of the absorption lines can give high transmission for a narrow range of frequencies.  

\begin{figure}[htb]
 \centerline{\includegraphics[width=11cm]{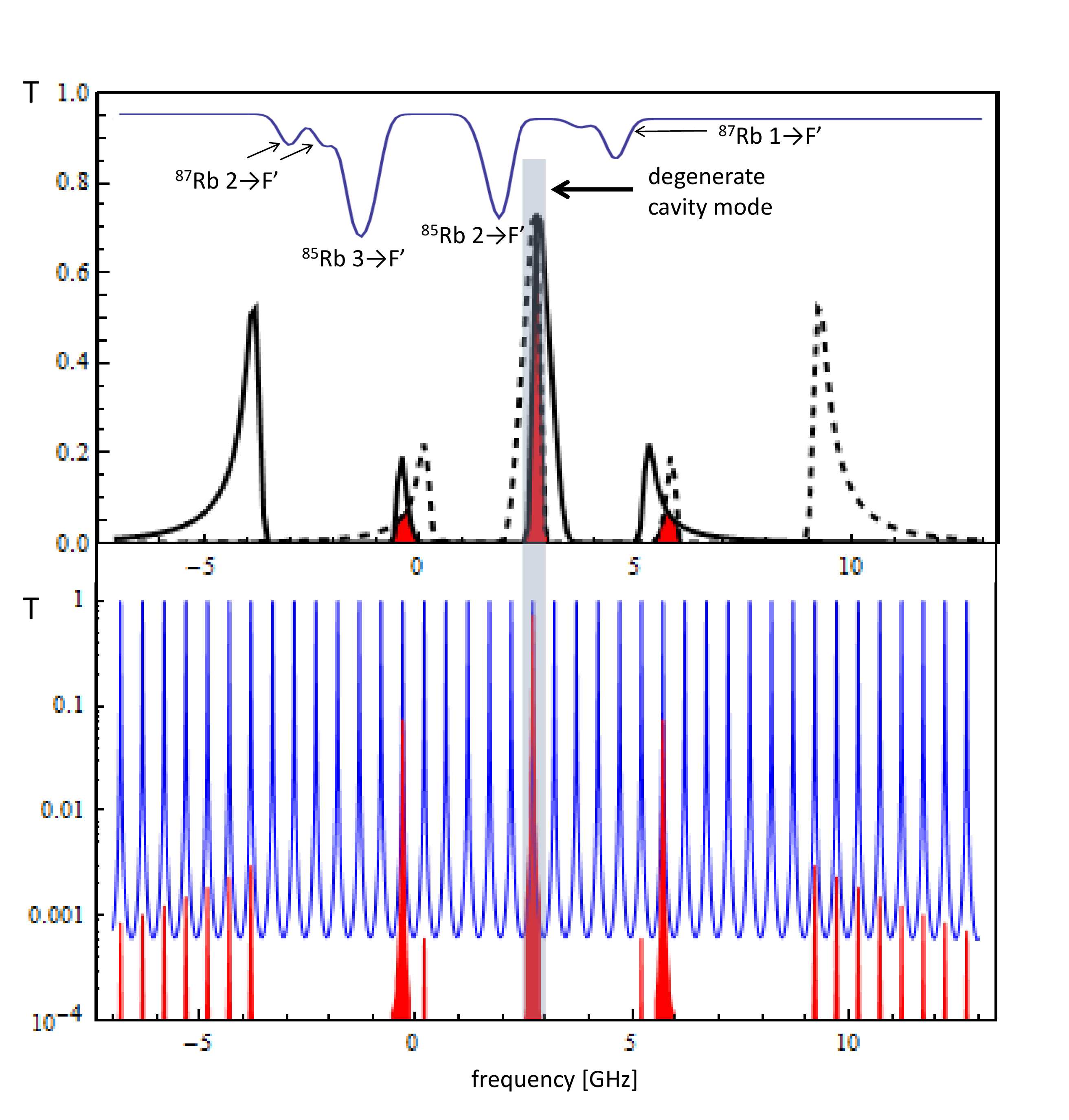}}
\caption{\label{fig:filterandcavity} Upper plot: reference transmission spectrum of room temperature natural abundance Rb (blue), filter spectrum (black) and a mirror filter spectrum with respect to the degenerate cavity mode (black dashed). Red shaded regions indicate transmission of correlated photon pairs. Lower plot: cavity output spectrum (blue) and FADOF-filtered cavity spectrum (red). The degenerate cavity mode coincides with the FADOF peak. Both figures have the same frequency scale.}
\end{figure}

In a previous work \cite{zielinskaOL2012} we demonstrated a FADOF filter on the D$_1$ line in Rb. The filter used in this experiment is the same as in the cited paper, except it has been modified to work for two orthogonal polarizations: instead of the crossed polarizers, we use a beam displacer before the cell, so that the two orthogonal polarizations travel along independent parallel paths in the cell.   After the cell we use a Wollaston prism to separate the near-resonant filtered light from the unrotated one. The optical axes of the two polarizing elements are oriented with precision mounts, and an extinction ratio of $1.8 \times 10^{-6}$ is reached. This strategy exploits the imaging capability of the filter.

Additionally, the setup has been supplemented with a half-waveplate placed before the Wollaston prism (HWP 2 in Fig. \ref{fig:opo}), which enables us to, in effect, turn on and off the filter. In the ``FADOF on'' condition, the waveplate axis is set parallel to the Wollaston axis (and thus the waveplate has no effect on the filter behaviour), the magnetic field is 4.5 mT and the temperature is 365 K.  In the ``FADOF off'' condition, no magnetic field is applied, the temperature of the cell is also 365 K and  HWP 2 is set to rotate the polarization by 90 degrees, in effect swapping the outputs, so that almost all the light is transmitted through the setup {without being filtered}. 

We optimized the filter using a common criterion for experiments with photon pairs:  we maximize the ratio of coincidences due to photon pairs belonging to the degenerate mode to coincidences due to other photon pairs. Because of energy conservation, the two photons in any SPDC pair will have frequencies symmetrically placed with respect to the degenerate mode; to prevent the pair from reaching the detectors, it suffices to block at least one of the photons.  In terms of filter performance, this means that it is possible to have near-perfect filtering even with transmission in some spectral windows away from the degenerate mode, provided the transmission is asymmetrical (Fig. \ref{fig:filterandcavity}).  Using this criterion we find the optimal conditions for the filter performance in our experiment to be 4.5 mT of magnetic field and the cell temperature of 365 K. The optimum filter performance requires the degenerate mode that should be filtered to coincide with the FADOF transmission peak at a fixed frequency (2.7 GHz to the red from the center of the Rb D$_1$ line).\\

\subsection{Detection}
\label{sec:detection}

The distribution of arrival times of photons in a Hanbury-Brown-Twiss configuration is useful to check that the filter effectively suppresses the non-degenerate modes of the type-I OPO described in the previous section. We collect the OPO output in a polarization maintaining fiber and send it through the filter setup. The filtered light is then coupled into balanced fiber beam splitters that send the photons to avalanche photo-detectors (APDs), connected to a time-of-flight analyzer (TOF) that allows us to measure the {second order correlation function} $G^{(2)}(T)$ (see Fig. \ref{fig:opo}). 

Since we are using single photon detectors, we need to reduce as much as possible the background due to stray light sources in the setup. The main source of background light is the counter-propagating beam that we inject in the OPO in order to lock the cavity length to be resonating at $\omega_0$. We tackle this problem using a chopped lock: the experiment switches at 85 Hz  between periods of data acquisition and periods of stabilization.  During {periods of} data acquisition, the AOM is off, and thus no locking beam is present.  During {periods of} stabilization, the AOM is on, and an electronic gate circuit is used to block electronic signals from the APDs, preventing recording of detections due to the locking beam photons. In addition, the polarization of the locking beam is orthogonal to that of the OPO output.

\newcommand{\ain}{a_{\rm in}}
\newcommand{\aout}{a_{\rm out}}
\newcommand{\bin}{b_{\rm in}}
\newcommand{\bout}{b_{\rm out}}

\subsection{Filter non-degenerate modes}

In this section we consider the second order correlation function of the field operators $\aout$ in a form:
\begin{equation}
   G^{(2)}(T) \propto \langle \aout^\dagger(t) \aout^\dagger(t+T) \aout(t+T) \aout(t) \rangle 
\end{equation}
for multimode (unfiltered) and single-mode (filtered) output of the OPO.

As shown  by Lu et al. \cite{lu2000}, $G^{(2)}(T)$ describing the output of a single-mode, far-below-threshold OPO has the form of double exponential decay
\begin{equation}
\label{g2single}
   G_{\rm single}^{(2)}(T)\propto e^{-|T|(\gamma_1+\gamma_2)},
\end{equation}
where the reflectivity of the output coupler is $r_1=\exp[{-\gamma_1 \tau}]$, the effective reflectivity resulting from intracavity losses is $r_2=\exp[{-\gamma_2 \tau}]$ and $\tau$ is the cavity round-trip time.  
An ideal narrowband filter would remove all the nondegenerate cavity-enhanced spontaneous down-conversion {CESPDC} 
 modes, reducing the $G^{(2)}(T)$  to $G_{\rm single}^{(2)}(T)$.  This filtering effect was demonstrated in \cite{WolfgrammPRL2011} for a type-II OPO and an induced dichroism atomic filter.

In \cite{lu2000} it is also predicted that when the filter is off, so that the output consists of $N$ cavity modes,  $G_{}^{(2)}(T)$ takes the form
\begin{eqnarray}
\label{eq:GmultiSin}
   G_{\rm multi}^{(2)}(T) &\propto & G_{\rm single}^{(2)}(T) \frac{\sin^2[(2N+1)\pi T/ \tau]}{(2N+1)\sin^2[\pi T/\tau]}
 \\ 
& \approx &  G_{\rm single}^{(2)}(T)\sum_{n=-\infty}^{\infty}\delta(T-n\tau),
\end{eqnarray}
i.e., with the same double exponential decay but modulated by a comb with a period equal to the cavity round-trip time $\tau$.  In our case the bandwidth of the output contains more than 200 cavity modes, and the fraction in Eq. (\ref{eq:GmultiSin}) is well approximated by a comb of Dirac delta functions. 

The comb period of $\tau = 1.99$~ns is comparable to the $t_{\rm bin} = 1$ ns resolution of our counting electronics, a digital time-of-flight counter (Fast ComTec P7888).  This counter assigns arrival times to the signal and idler arrivals relative to an internal clock.  We take the ``window function'' for the $i$th bin, i.e., the probability of an arrival at time $T$ being assigned to that bin,  to be
\begin{equation}
   f^{(i)}(T)=
\begin{cases}
    1,& \text{if } T\in [i t_{bin}, (i+1) t_{bin}] \,,   \\
    0,              & \text{otherwise}\,.
\end{cases}
\end{equation}
 
Without loss of generality we assign the signal photon's bin as $i=0$, and we include an unknown relative delay $T_0$ between signal and idler due to path length, electronics, cabling, and so forth.  For a given signal arrival time $t_s$, the rate of idler arrivals in the $i$th bin is $\int dt_i \,  f^{(i)}(t_i) G_{\rm multi}^{(2)}(t_i - t_s - T_0)$ ($t_i$ is the idler arrival time).  This expression must be averaged over the possible $t_s$ within bin $i=0$.  We also include the ``accidental'' coincidence rate $G_{\rm acc}^{(2)}= t_{\rm bin} R_1 R_2$, where $R_1, R_2$ are the singles detection rates at detectors $1,2$, respectively. The rate at which coincidence events are registered with $i$ bins of separation is  then 
\begin{eqnarray}
   G_{\rm multi,det}^{(2)}(i) &= & \frac{1}{t_{\rm bin}} \int  dt_s \, f^{(0)}(t_s) \int dt_i \,  f^{(i)}(t_i) G_{\rm multi}^{(2)}(t_i - t_s - T_0) +  G_{\rm acc}^{(2)} \\
   &=&  \sum_{n=-\infty}^{\infty} G_{\rm single}^{(2)}(n \tau) \frac{1}{t_{\rm bin}} \int_0^{t_{\rm bin}} dt_s \,  f^{(i)}(t_s + T_0 + n \tau)  +  G_{\rm acc}^{(2)}. 
\end{eqnarray}

 We take $T_0$ is a free parameter in fitting to the data. Note that if we write $T_0=k t_{\rm bin}+\delta$ then the simultaneous events fall into $k$th bin and $\delta\in[-t_{\rm bin}/2,t_{\rm bin}/2]$ determines where the histogram has the maximum visibility due to the beating between the 1 ns sampling frequency of the detection system and the 1.99 ns comb period.   APD time resolution is estimated to be $350$ ps FWHM (manufacturer's specification), i.e. significantly less than the TOF uncertainty, and is not included here.

\begin{figure}
\centering
  \includegraphics[width= 0.9\textwidth]{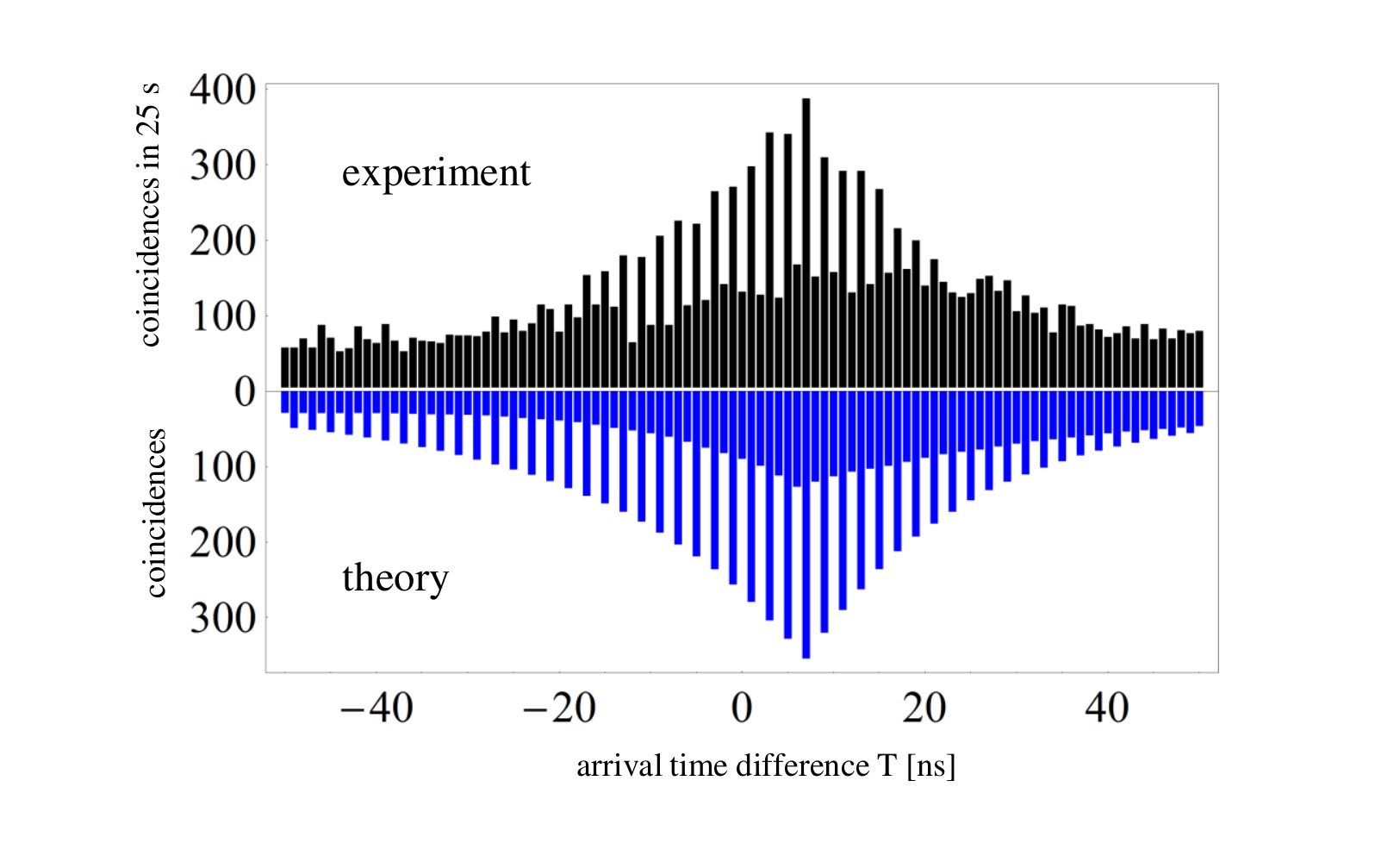}

\caption{\label{fig:fadofoff} Histograms of arrival time differences for FADOF off compared to theoretical model (both include the background due to accidental coincidences and the artefacts resulting from 1 ns resolution of the counting electronics).  
}
\end{figure}

\begin{figure}
\centering

  \includegraphics[width=0.9 \textwidth]{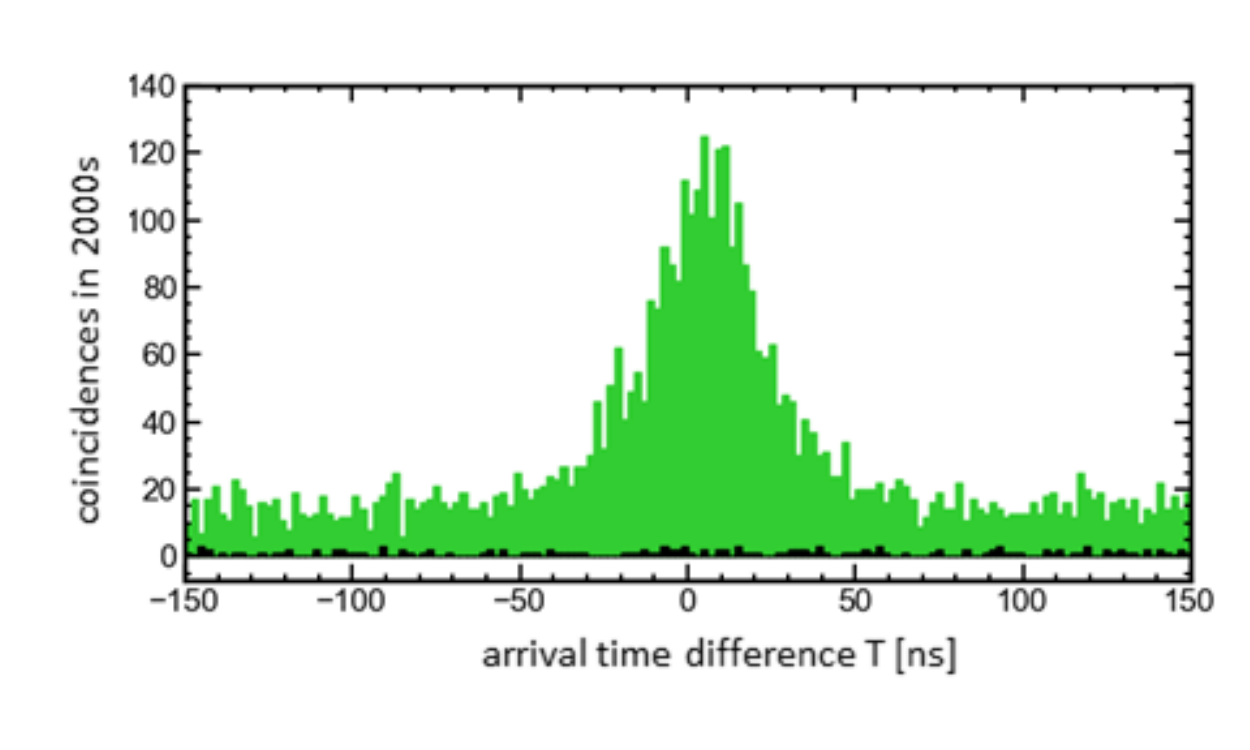}
\caption{\label{fig:fadofon} Histograms of the differences of arrival times of the photon pairs for FADOF on (green) and FADOF on with hot cell on the path (black). No background has been subtracted.
}
\end{figure}

Histograms of photon arrival time differences for the ``FADOF off'' and ``FADOF on'' configurations are shown in Figs. \ref{fig:fadofoff} and \ref{fig:fadofon}, respectively.
We observe that the double exponential full-width at half-maximum (FWHM) corresponds to the predicted one of 26~ns. Moreover, we notice how the comb structure is not present in the ``filter on'' data, as expected if the filter blocks all pairs not in the degenerate mode.  

\subsection{Spectral purity}

\newcommand{\SP}{P_{\rm S}}

According to the theoretical filter spectrum from \cite{zielinskaOL2012}, we estimate that 98\% of the atom-resonant photon pairs come from degenerate mode (see Fig.~\ref{fig:filterandcavity}). In order to test how much light outside the Rubidium D$_1$ line can pass through our FADOF, we split the light equally between the two different polarization paths of the filter setup by means of a half-wave plate put before the beam displacer (HWP 1 in Fig.\ref{fig:opo}). A natural-abundance Rb vapor cell, with 10~Torr of  N$_2$ buffer gas and heated until it is opaque for resonant light, is inserted in one of the paths  after the filter.  The collisionally-broadened absorption from this cell blocks the entire FADOF transmission window, allowing us to compare the arrival time histograms with and without the resonant component. 

The number of photons detected after passing through the hot Rb cell is comparable to dark counts, meaning that most of the filtered light is at the chosen frequency $\omega_0$.  We define the spectral purity $\SP$ of the FADOF as $\SP \equiv 1 - c_{HC}/c_{F}$, where $c_{HC}$ ($c_{F}$) is the number of photon pairs which were recorded within a coincidence windows of 50~ns in the path with (without) the hot cell. Considering raw coincidences (no background subtraction), we obtain $\SP = 0.98$, meaning that the filtered signal is remarkably pure, as only the $2\%$ of the recorded pairs are out of the filter spectrum.  This $2\%$ agrees with measurements of the polarization extinction ratio with the FADOF off, i.e., it is due to technical limitations of the polarization optics and could in principle be improved. Knowing that $98\%$ of the photon pairs transmitted through the filter within the Rb resonance come from the degenerate cavity mode (due to filter spectrum), we conclude that  $96\%$ of the pairs exiting the filter come from the degenerate mode.

\section{Continuous-variable measurement}

In this section we describe a noise contribution that in principle the FADOF filter might add to the filtered beam. Since the filter is a passive, linear device, the transformation that the anihilation operator undergoes when passing the filter is unitary:

\begin{eqnarray}
a_{\rm out}\rightarrow t a+r b
\end{eqnarray}
where $r^*t+rt^*=0$, $|r|^2+|t|^2=1$, operator $a$ represents the probe field and operator $b$ the vacuum field. Let us assume that filter transmission $t = \bar{t} + \delta t$ randomly fluctuates around mean value $\bar{t}$ with an amplitude $\delta t$. 

In order to estimate the effect such a device would have on the probe beam, we calculate the variance of the detected quadrature operator $\hat{X}_\theta=a \exp[i \theta]+a^\dagger \exp[-i\theta]$ averaged over the angle $\theta$ on the input state being a mixture of coherent states $\rho=\int d^2\alpha P(\alpha)|\alpha\rangle\langle\alpha|$ with a mean value $\bar{\alpha}$ fluctuating with an amplitude $\delta \alpha$. We find the quadrature noise has a form

\begin{eqnarray}
\langle{\rm var} (\hat{X}_{\theta})\rangle_\theta=1+2{\rm Re}(\bar{t}\bar{\alpha}){\rm Re}(\bar{\alpha}\delta t +\bar{t}\delta \alpha) +2{\rm Im} (\bar{t}\bar{\alpha}){\rm Im}(\bar{\alpha}\delta t + \bar{t}\delta \alpha)+ O(\delta \alpha \delta t)
\label{eq:QuadNoise}
\end{eqnarray}

Attenuation of the input probe intensity by a factor of $T_{\rm ND}$ with a neutral-density filter effects the changes $t\rightarrow t$, $\alpha \rightarrow t_{\rm ND} \alpha$, $\delta\alpha \rightarrow  {t_{\rm ND}} \delta\alpha$, where $t_{\rm ND} \equiv \sqrt{T_{\rm ND}}$ is the amplitude transmission.  Scaling with $t_{\rm ND}$ allows us to separate the different contributions:  The first term is the SQL, and scales as  $t_{\rm ND}^0$.  
The second term is noise introduced by the filter, and scales as $t_{\rm ND}^{2}$. The last term vanishes if $\delta \alpha$ and $\delta t$ are uncorrelated, and even without this assumption can be assumed much smaller than the other terms.  The average signal $\langle\hat{X}_{\rm out} \rangle$ scales as $ t_{\rm ND}$, providing a convenient measure of
the input power.  We perform an experiment in order to estimate the filter technical noise at the probe power similar to the intensity of non-classical light from our source.

\begin{figure}
\centering
\includegraphics[width=0.8\textwidth]{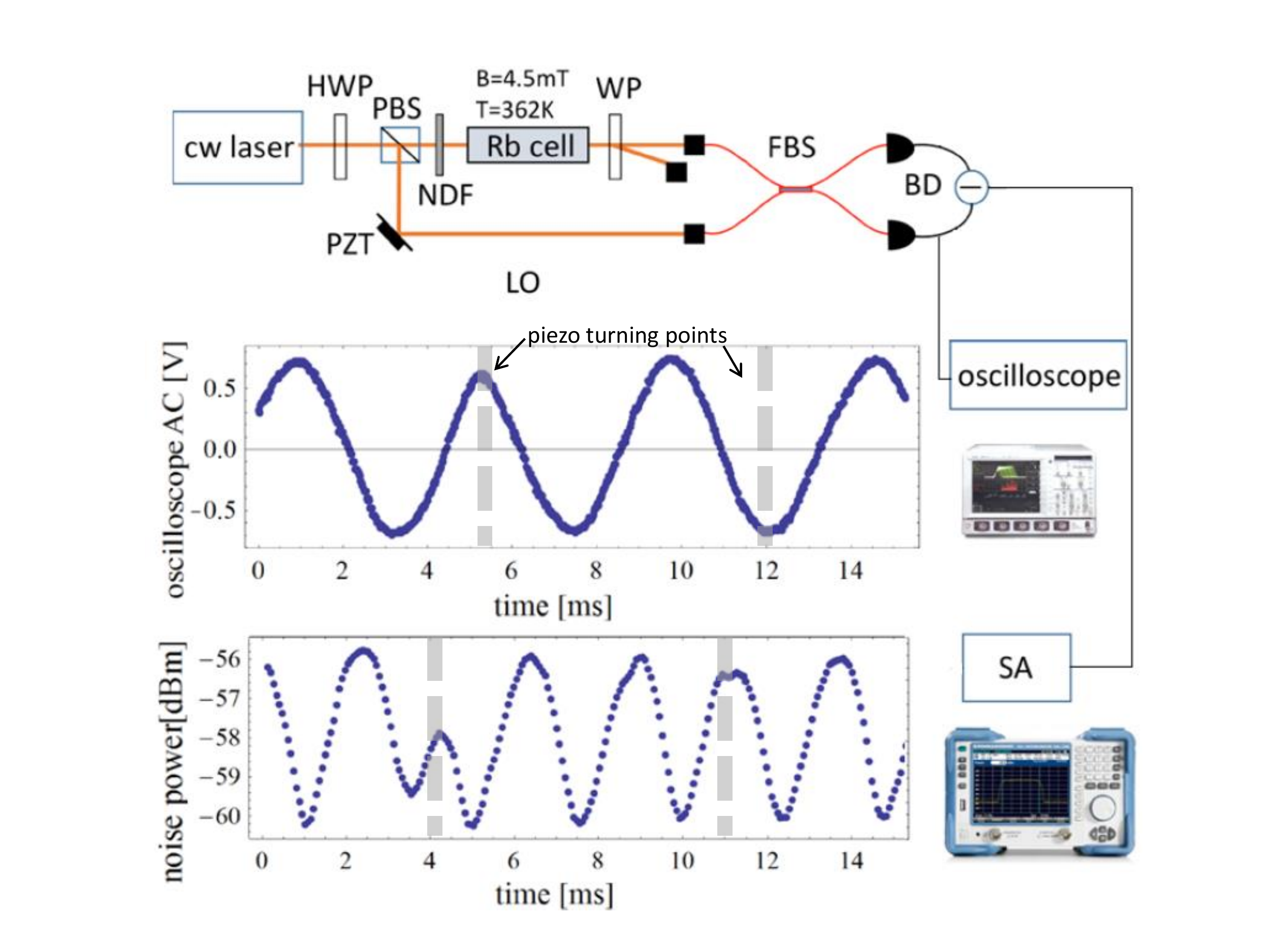}
\caption{\label{fig:fadofnoise} Experimental setup and traces from the oscilloscope and the spectrum analyzer. HWP- half-wave plate, PBS- polarization beam-splitter, WP - Wollaston prism, FBS- 50/50 fiber beam-splitter, BD - balanced detector, PZT - piezoelectric actuator, SA - spectrum analyzer. }
\end{figure}
The experimental setup is shown in Fig. \ref{fig:fadofnoise}. A continuous wave laser, stabilized at the FADOF peak frequency (as described in the section \ref{sec:Experiment}) is split into a strong (1 mW) local oscillator beam (LO) and a weak (1 $\mu$W) probe beam passing through the FADOF. The relative phase $\theta$ of the two beams is controlled by a mirror mounted on the piezoelectric actuator driven with a triangle wave at approximately 70 Hz. The two beams are coupled into single-mode fibers and combined on a fiber beamsplitter, the outputs of which are fed to the balanced detector (Thorlabs PDB450A) with a gain of $10^5$ over a bandwidth of 4.5 MHz. The difference output of the balanced detector is recorded by a spectrum analyzer in a zero-span mode with center frequency of 2 MHz, resolution bandwidth of 300 kHz and video bandwidth of 100 Hz. The monitor output of  one of the two photodiodes comprising the balanced detector is simultaneously recorded on an AC-coupled oscilloscope (SC). 

As shown in Fig. \ref{fig:fadofnoise}, we observe oscillations in both the SA and SC signals versus $\theta$.   The SC signal indicates the mean detected quadrature.  We keep the peak-to-peak variation as a convenient measure of the field strength.  The SA signal indicates the noise of the detected quadrature.  This oscillates with $\theta$, presumably because of excess laser phase noise leading to extra variance in the phase quadrature.  As seen in the figure, it oscillates at twice the rate of the SC oscillations, as expected for a noise measurement.  

\begin{figure}
\centering
\includegraphics[width=0.7\textwidth]{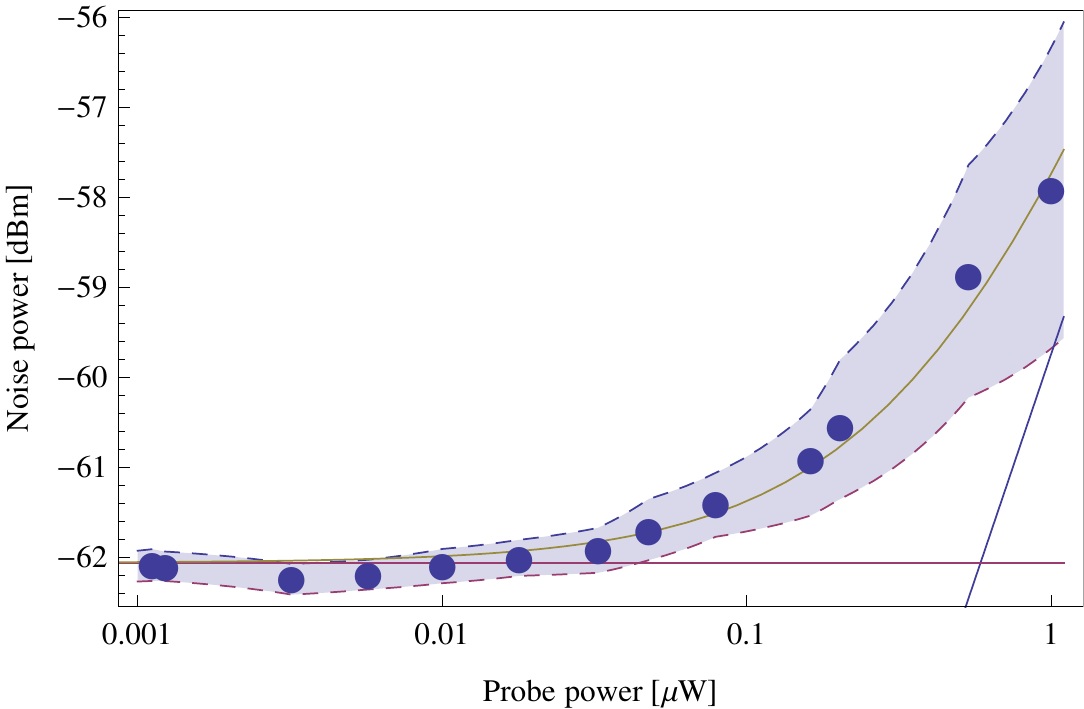}
\caption{\label{fig:fadofnoiseplot} Average maxima and minima (dashed lines) and mean value (blue circles) of the noise oscillations detected on the SA due to the laser phase noise, as a function of the variance of the oscilloscope signal proportional to the probe power. The solid lines represent the fitted noise model (brown), a sum of a term linear with power (blue) and shot noise (red).
 }
\end{figure}

In Fig. \ref{fig:fadofnoiseplot} we plot the mean, and the maximum and minimum, averaged over a few cycles, of the noise oscillations on the SA as a function of the probe power (proportional to the variance of the oscilloscope signal).  The measurement runs from zero probe power to 1 $\mu$W.  The power at  1 $\mu$W is measured with a power meter, which provides a calibration for the SC measurements.  As seen in the figure, below about 10 nW of probe the mean noise level drops to the shot noise level and the oscillations disappear. 10 nW corresponds to a photon flux of $4 \times 10^{10}$ photons per second, much larger than typical photon fluxes for our OPO with 8 MHz bandwidth, e.g. $\approx 10^7$ photons/s at 3 dB of squeezing.  A dependence of mean noise power on the probe power (constant and linear term) is fitted according to the model of Eq. (\ref{eq:QuadNoise}) and represented in the Fig. \ref{fig:fadofnoiseplot}. Extrapolating the contribution of noise from the filter (linear term) we estimate that at the power level of $10^7$ photons/s, the filter would introduce approximately $-150$ dBm of electronic noise, corresponding to $-88$ dB of noise with respect to the shot noise level.

The measurement we performed shows that the FADOF does not add significant amount of noise to the coherent probe, which in turn indicates that it is possible to send through a squeezed state without destroying it. For example, input squeezing of 6 dB, after passing through the filter would be reduced to 3.2 dB due to the filter's 70$\%$ transmission.  

\section{Conclusion}

We have demonstrated the use of a high-performance atomic filter to separate fully degenerate photon pairs from the broadband emission of a sub-threshold OPO, or equivalently a CESPDC source.  The filter, based on the FADOF principle, achieves simultaneously sufficient out-of-band rejection to allow accurate photon-counting detection and sufficient transmission to preserve continuous-variable characteristics such as squeezing.  Combining these properties in the narrow-band regime is critical to generation of hybrid continuous-variable/discrete-variable states compatible with atomic systems, e.g. quantum memories.  The results may also advance proposals for loophole-free Bell tests,  quantum metrology, and quantum computing.  

\section*{Acknowledgement}
We thank Wilhelm Kiefer and Ilja Gerhardt for an improved refractive index calculator.  This work was supported by the Spanish MINECO project MAGO (Ref.
FIS2011-23520), European Research Council project AQUMET and by Fundaci\'{o} Privada CELLEX. J.Z. was supported by the
FI-DGR PhD-fellowship program of the Generalitat of Catalonia.

\end{document}